\pdfoutput=1

\documentclass[12pt,a4paper]{article}

\usepackage{ifthen} 
\newboolean{pdflatex}
\setboolean{pdflatex}{true} 

\newboolean{articletitles}
\setboolean{articletitles}{true} 

\newboolean{uprightparticles}
\setboolean{uprightparticles}{false} 

\def\paperauthors{LHCb collaboration} 
\def\paperasciititle{First observation of the semileptonic decay B+->ppbarmunu} 
\def\papertitle{Observation of the semileptonic decay $\Bp\to\ppbar\mup\neum$} 
\def\paperkeywords{{High Energy Physics}, {LHCb}} 
\def\papercopyright{\the\year\ CERN for the benefit of the LHCb collaboration} 
\def\paperlicence{CC-BY-4.0 licence}
\def\paperlicenceurl{https://creativecommons.org/licenses/by/4.0/}

\usepackage[top=1in, bottom=1.25in, left=1in, right=1in]{geometry}

\usepackage{microtype}
\usepackage{lineno}  
\usepackage{xspace} 
\usepackage{caption} 

\usepackage{graphicx}  
\usepackage{color}
\usepackage{colortbl}
\usepackage{multirow}
\usepackage{booktabs}
\usepackage[percent]{overpic}
\graphicspath{{./figs/}} 
\DeclareGraphicsExtensions{.pdf,.PDF,png,.PNG}

\usepackage{amsmath} 
\usepackage{amssymb}
\usepackage{amsfonts}
\usepackage{upgreek} 

\newcommand*\patchAmsMathEnvironmentForLineno[1]{%
\expandafter\let\csname old#1\expandafter\endcsname\csname #1\endcsname
\expandafter\let\csname oldend#1\expandafter\endcsname\csname
end#1\endcsname
 \renewenvironment{#1}%
   {\linenomath\csname old#1\endcsname}%
   {\csname oldend#1\endcsname\endlinenomath}%
}
\newcommand*\patchBothAmsMathEnvironmentsForLineno[1]{%
  \patchAmsMathEnvironmentForLineno{#1}%
  \patchAmsMathEnvironmentForLineno{#1*}%
}
\AtBeginDocument{%
\patchBothAmsMathEnvironmentsForLineno{equation}%
\patchBothAmsMathEnvironmentsForLineno{align}%
\patchBothAmsMathEnvironmentsForLineno{flalign}%
\patchBothAmsMathEnvironmentsForLineno{alignat}%
\patchBothAmsMathEnvironmentsForLineno{gather}%
\patchBothAmsMathEnvironmentsForLineno{multline}%
\patchBothAmsMathEnvironmentsForLineno{eqnarray}%
}

\usepackage{hyperxmp}

\usepackage[pdftex,
            pdfauthor={\paperauthors},
            pdftitle={\paperasciititle},
            pdfkeywords={\paperkeywords},
            pdfcopyright={Copyright (C) \papercopyright},
            pdflicenseurl={\paperlicenceurl}]{hyperref}

\usepackage[colorinlistoftodos,textsize=scriptsize]{todonotes}

\usepackage[all]{hypcap} 

\usepackage{xspace} 
\usepackage{upgreek}

\def\ppbar {\ensuremath{\proton\antiproton}{}\xspace}
\def\btoppbarmunu{\mbox{\ensuremath{\Bp\to\proton\antiproton\mup\neum}{}}\xspace}
\def\bptoppbarmunu{\mbox{\ensuremath{\Bp\to\proton\antiproton\mup\neum}{}}\xspace}
\def\bptoppbarlnu{\mbox{\ensuremath{\Bp\to\proton\antiproton \ellp\nu_{\ell}}{}}\xspace}
\def\mppbar{\ensuremath{m(\proton\antiproton)}{}\xspace}
\def\mcorr{\ensuremath{m_{\mathrm{corr}}}{}\xspace}

\def\bptojpsikp{\mbox{\ensuremath{\Bp\to\jpsi\Kp}{}}\xspace}

\def\btoyyx{\mbox{\ensuremath{B\to Y\overline{Y}'X}{}}\xspace}

\def\lhcb   {\mbox{LHCb}\xspace}

\def\belle  {\mbox{Belle}\xspace}

\def\MagUp {\mbox{\em Mag\kern -0.05em Up}\xspace}

\ifthenelse{\boolean{uprightparticles}}%
{

 \def\Peta        {\ensuremath{\upeta}\xspace}

 \def\Pmu         {\ensuremath{\upmu}\xspace}                 
 \def\Pnu         {\ensuremath{\upnu}\xspace}                 
                  
 \def\Ppi         {\ensuremath{\uppi}\xspace}

 \def\Ppsi        {\ensuremath{\uppsi}\xspace}

 \def\PDelta      {\ensuremath{\Delta}\xspace}                 
 \def\PXi         {\ensuremath{\Xi}\xspace}                 
 \def\PLambda     {\ensuremath{\Lambda}\xspace}                 
 \def\PSigma      {\ensuremath{\Sigma}\xspace}                 
 \def\POmega      {\ensuremath{\Omega}\xspace}                 
 \def\PUpsilon    {\ensuremath{\Upsilon}\xspace}

 \def\PB      {\ensuremath{\mathrm{B}}\xspace}                 
                  
 \def\PD      {\ensuremath{\mathrm{D}}\xspace}

 \def\PJ      {\ensuremath{\mathrm{J}}\xspace}                 
 \def\PK      {\ensuremath{\mathrm{K}}\xspace}

 \def\Pb      {\ensuremath{\mathrm{b}}\xspace}                 
 \def\Pc      {\ensuremath{\mathrm{c}}\xspace}                 
                  
 \def\Pe      {\ensuremath{\mathrm{e}}\xspace}

 \def\Pi      {\ensuremath{\mathrm{i}}\xspace}

 \def\Pp      {\ensuremath{\mathrm{p}}\xspace}

 \def\Ps      {\ensuremath{\mathrm{s}}\xspace}                 
                  
 \def\Pu      {\ensuremath{\mathrm{u}}\xspace}

 \def\thebaroffset{0.0em}
}
{

 \def\Peta        {\ensuremath{\eta}\xspace}

 \def\Pmu         {\ensuremath{\mu}\xspace}                 
 \def\Pnu         {\ensuremath{\nu}\xspace}                 
                  
 \def\Ppi         {\ensuremath{\pi}\xspace}

 \def\Ppsi        {\ensuremath{\psi}\xspace}                 
                  
 \mathchardef\PDelta="7101
 \mathchardef\PXi="7104
 \mathchardef\PLambda="7103
 \mathchardef\PSigma="7106
 \mathchardef\POmega="710A
 \mathchardef\PUpsilon="7107
                  
 \def\PB      {\ensuremath{B}\xspace}                 
                  
 \def\PD      {\ensuremath{D}\xspace}

 \def\PJ      {\ensuremath{J}\xspace}                 
 \def\PK      {\ensuremath{K}\xspace}

 \def\Pb      {\ensuremath{b}\xspace}                 
 \def\Pc      {\ensuremath{c}\xspace}                 
                  
 \def\Pe      {\ensuremath{e}\xspace}

 \def\Pi      {\ensuremath{i}\xspace}

 \def\Pp      {\ensuremath{p}\xspace}

 \def\Ps      {\ensuremath{s}\xspace}                 
                  
 \def\Pu      {\ensuremath{u}\xspace}

 \def\thebaroffset{0.18em}
}
\newcommand{\offsetoverline}[2][\thebaroffset]{\kern #1\overline{\kern -#1 #2}}%

\makeatletter
\ifcase \@ptsize \relax
  \newcommand{\miniscule}{\@setfontsize\miniscule{4}{5}}
\or
  \newcommand{\miniscule}{\@setfontsize\miniscule{5}{6}}
\or
  \newcommand{\miniscule}{\@setfontsize\miniscule{5}{6}}
\fi
\makeatother

\DeclareRobustCommand{\optbar}[1]{\shortstack{{\miniscule (\rule[.5ex]{1.25em}{.18mm})}
  \\ [-.7ex] $#1$}}

\def\ep         {{\ensuremath{\Pe^+}}\xspace}

\def\mup        {{\ensuremath{\Pmu^+}}\xspace}
\def\mun        {{\ensuremath{\Pmu^-}}\xspace} 

\def\ellp       {{\ensuremath{\ell^+}}\xspace}

\def\neu        {{\ensuremath{\Pnu}}\xspace}

\def\neue       {{\ensuremath{\neu_e}}\xspace}

\def\neum       {{\ensuremath{\neu_\mu}}\xspace}

\def\uquark    {{\ensuremath{\Pu}}\xspace}

\def\squark    {{\ensuremath{\Ps}}\xspace}

\def\cquark    {{\ensuremath{\Pc}}\xspace}

\def\bquark    {{\ensuremath{\Pb}}\xspace}

\def\pion   {{\ensuremath{\Ppi}}\xspace}

\def\pip    {{\ensuremath{\pion^+}}\xspace}
\def\pim    {{\ensuremath{\pion^-}}\xspace}

\def\kaon    {{\ensuremath{\PK}}\xspace}

\def\KorKbar {\kern \thebaroffset\optbar{\kern -\thebaroffset \PK}{}\xspace}

\def\Kp      {{\ensuremath{\kaon^+}}\xspace}

\def\Dbar    {{\ensuremath{\offsetoverline{\PD}}}\xspace}
\def\D       {{\ensuremath{\PD}}\xspace}

\def\DorDbar {\kern \thebaroffset\optbar{\kern -\thebaroffset \PD}\xspace}

\def\Dzb     {{\ensuremath{\Dbar{}^0}}\xspace}

\def\Dm      {{\ensuremath{\D^-}}\xspace}

\def\Dstarm  {{\ensuremath{\D^{*-}}}\xspace}

\def\B       {{\ensuremath{\PB}}\xspace}

\def\BorBbar {\kern \thebaroffset\optbar{\kern -\thebaroffset \PB}\xspace}
\def\Bz      {{\ensuremath{\B^0}}\xspace}

\def\Bd      {{\ensuremath{\B^0}}\xspace}

\def\BdorBdbar {\kern \thebaroffset\optbar{\kern -\thebaroffset \Bd}\xspace}
\def\Bu      {{\ensuremath{\B^+}}\xspace}

\def\Bp      {{\ensuremath{\Bu}}\xspace}

\def\Bs      {{\ensuremath{\B^0_\squark}}\xspace}

\def\BsorBsbar {\kern \thebaroffset\optbar{\kern -\thebaroffset \Bs}\xspace}

\def\jpsi     {{\ensuremath{{\PJ\mskip -3mu/\mskip -2mu\Ppsi\mskip 2mu}}}\xspace}
\def\psitwos  {{\ensuremath{\Ppsi{(2S)}}}\xspace}

\def\etac     {{\ensuremath{\Peta_\cquark}}\xspace}

\def\Y#1S{\ensuremath{\PUpsilon{(#1S)}}\xspace}

\def\proton      {{\ensuremath{\Pp}}\xspace}
\def\antiproton  {{\ensuremath{\overline \proton}}\xspace}

\def\Deltares    {{\ensuremath{\PDelta}}\xspace}

\def\Lz          {{\ensuremath{\PLambda}}\xspace}
\def\Lbar        {{\ensuremath{\offsetoverline{\PLambda}}}\xspace}
\def\LorLbar     {\kern \thebaroffset\optbar{\kern -\thebaroffset \PLambda}\xspace}

\def\Lcbar       {{\ensuremath{\Lbar{}^-_\cquark}}\xspace}

\def\to                 {\ensuremath{\rightarrow}\xspace}

\def\AT#1     {\ensuremath{A_{\mathrm{T}}^{#1}}\xspace}           

\def\C#1      {\ensuremath{\mathcal{C}_{#1}}\xspace}                       
\def\Cp#1     {\ensuremath{\mathcal{C}_{#1}^{'}}\xspace}                    
\def\Ceff#1   {\ensuremath{\mathcal{C}_{#1}^{\mathrm{(eff)}}}\xspace}        
\def\Cpeff#1  {\ensuremath{\mathcal{C}_{#1}^{'\mathrm{(eff)}}}\xspace}       
\def\Ope#1    {\ensuremath{\mathcal{O}_{#1}}\xspace}                       
\def\Opep#1   {\ensuremath{\mathcal{O}_{#1}^{'}}\xspace}                    

\newcommand{\nospaceunit}[1]{\ensuremath{\text{#1}}}       
\newcommand{\aunit}[1]{\ensuremath{\text{\,#1}}}       

\newcommand{\tev}{\aunit{Te\kern -0.1em V}\xspace}
\newcommand{\gev}{\aunit{Ge\kern -0.1em V}\xspace}
\newcommand{\mev}{\aunit{Me\kern -0.1em V}\xspace}
\newcommand{\kev}{\aunit{ke\kern -0.1em V}\xspace}
\newcommand{\ev}{\aunit{e\kern -0.1em V}\xspace}
\newcommand{\mevc}{\ensuremath{\aunit{Me\kern -0.1em V\!/}c}\xspace}
\newcommand{\gevc}{\ensuremath{\aunit{Ge\kern -0.1em V\!/}c}\xspace}
\newcommand{\mevcc}{\ensuremath{\aunit{Me\kern -0.1em V\!/}c^2}\xspace}
\newcommand{\gevcc}{\ensuremath{\aunit{Ge\kern -0.1em V\!/}c^2}\xspace}

\def\mum  {\ensuremath{\,\upmu\nospaceunit{m}}\xspace}

\def\fb   {\ensuremath{\aunit{fb}}\xspace}
\def\invfb   {\ensuremath{\fb^{-1}}\xspace}

\def\gsim{{~\raise.15em\hbox{$>$}\kern-.85em
          \lower.35em\hbox{$\sim$}~}\xspace}
\def\lsim{{~\raise.15em\hbox{$<$}\kern-.85em
          \lower.35em\hbox{$\sim$}~}\xspace}

\def\pt         {\ensuremath{p_{\mathrm{T}}}\xspace}

\def\ptot       {\ensuremath{p}\xspace}

\def\evtgen     {\mbox{\textsc{EvtGen}}\xspace}

\def\geant      {\mbox{\textsc{Geant4}}\xspace}

\def\photos     {\mbox{\textsc{Photos}}\xspace}

\def\pythia     {\mbox{\textsc{Pythia}}\xspace}

\def\tell1  {TELL1\xspace}
\def\ukl1   {UKL1\xspace}

\newcommand{\etc}{\mbox{\itshape etc.}\xspace}

\usepackage{cite} 
\usepackage{mciteplus}

\usepackage{tikz}
\usetikzlibrary{math}
\usepackage{siunitx}

\tikzmath{
    \staterrone = (1.9577213129+1.9443150546)/(2.*21.3687307111);
    \staterrtwo = (1.139313305+1.2287300171)/(2.*21.3949604176);
    \staterrthree = (0.7980458186+0.7692187491)/(2.*6.286059659);
    \staterrfour = (0.4625062629+0.4115647666)/(2.*1.7246551547);
    \staterrfive = (0.0270617035+0.0265553004)/(2.*0.0899782231);
    \staterronepc = 100.*(1.9577213129+1.9443150546)/(2.*21.3687307111);
    \staterrtwopc = 100.*(1.139313305+1.2287300171)/(2.*21.3949604176);
    \staterrthreepc = 100.*(0.7980458186+0.7692187491)/(2.*6.286059659);
    \staterrfourpc = 100.*(0.4625062629+0.4115647666)/(2.*1.7246551547);
    \staterrfivepc = 100.*(0.0270617035+0.0265553004)/(2.*0.0899782231);
    \biasone =  2.3*\staterrone;
    \biastwo = 2.4*\staterrtwo;
    \biasthree = 7.6*\staterrthree;
    \biasfour = 9.7*\staterrfour; 
    \biasfive = 26.3*\staterrfive;
    \syssigone = 6.5*\staterrone;
    \syssigtwo = 36.5*\staterrtwo;
    \syssigthree = 23.8*\staterrthree;
    \syssigfour = 19.0*\staterrfour;
    \syssigfive = 33.3*\staterrfive;
    \sysbootstrapone = 3.5*\staterrone; 
    \sysbootstraptwo = 0.4*\staterrtwo;  
    \sysbootstrapthree = 2.5*\staterrthree; 
    \sysbootstrapfour = 9.4*\staterrfour; 
    \sysbootstrapfive = 17.4*\staterrfive; 
    \syssmoothone = 0.2*\staterrone;
    \syssmoothtwo = 20.3*\staterrtwo;  
    \syssmooththree = 21.3*\staterrthree; 
    \syssmoothfour = 31.2*\staterrfour; 
    \syssmoothfive = 11.6*\staterrfive; 
    \sysmisidone = 10.0*\staterrone;  
    \sysmisidtwo = 2.3*\staterrtwo;
    \sysmisidthree = 4.5*\staterrthree; 
    \sysmisidfour = 20.6*\staterrfour;
    \sysmisidfive = 45.4*\staterrfive; 
    \syscombione = 9.9*\staterrone;
    \syscombitwo = 21.2*\staterrtwo;
    \syscombithree = 9.7*\staterrthree;
    \syscombifour = 33.6*\staterrfour;
    \syscombifive = 15.7*\staterrfive;
    \systotone = sqrt(\biasone*\biasone + \syssigone*\syssigone + \sysbootstrapone*\sysbootstrapone +\syssmoothone*\syssmoothone + \sysmisidone*\sysmisidone + \syscombione*\syscombione)*\staterrone;
    \systottwo = sqrt(\biastwo*\biastwo + \syssigtwo*\syssigtwo + \sysbootstraptwo*\sysbootstraptwo +\syssmoothtwo*\syssmoothtwo + \sysmisidtwo*\sysmisidtwo + \syscombitwo*\syscombitwo)*\staterrtwo;
    \systotthree = sqrt(\biasthree*\biasthree + \syssigthree*\syssigthree + \sysbootstrapthree*\sysbootstrapthree +\syssmooththree*\syssmooththree + \sysmisidthree*\sysmisidthree + \syscombithree*\syscombithree)*\staterrthree;
    \systotfour = sqrt(\biasfour*\biasfour + \syssigfour*\syssigfour + \sysbootstrapfour*\sysbootstrapfour +\syssmoothfour*\syssmoothfour + \sysmisidfour*\sysmisidfour + \syscombifour*\syscombifour)*\staterrfour;
    \systotfive = sqrt(\biasfive*\biasfive + \syssigfive*\syssigfive + \sysbootstrapfive*\sysbootstrapfive +\syssmoothfive*\syssmoothfive + \sysmisidfive*\sysmisidfive + \syscombifive*\syscombifive)*\staterrfive;
    \effsystkrone = 0.693;
    \effsystkronert = 0.213;
    \effsystkroneTOT = 0.27973148145559334;
    \effsystkrtwo = 0.557;
    \effsystkrtwort = 0.852;
    \effsystkrtwoTOT = 0.5734627484523329;
    \effsystkrthree = 0.394;
    \effsystkrthreert = 0.864;
    \effsystkrthreeTOT = 0.5629197160974765;
    \effsystkrfour = 0.496;
    \effsystkrfourrt = 0.59;
    \effsystkrfourTOT = 0.4157131737285465;
    \effsystkrfive = 0.388;
    \effsystkrfivert = 0.458;
    \effsystkrfiveTOT = 0.32191202942341485;
    \effsystsimstatone = 4.799;
    \effsystsimstatonert = 4.961;
    \effsystsimstatoneTOT = 3.633117287710352;
    \effsystsimstattwo = 4.418;
    \effsystsimstattwort = 4.332;
    \effsystsimstattwoTOT = 3.178375953996027;
    \effsystsimstatthree = 4.389;
    \effsystsimstatthreert = 4.289 ;
    \effsystsimstatthreeTOT = 3.1510687560637614;
    \effsystsimstatfour = 4.382;
    \effsystsimstatfourrt = 4.277;
    \effsystsimstatfourTOT = 3.1494991243656094;
    \effsystsimstatfive = 4.23;
    \effsystsimstatfivert = 4.028;
    \effsystsimstatfiveTOT = 2.9784250621007136;
    \effsysttrackingone = 2.738;
    \effsysttrackingonert = 2.739;
    \effsysttrackingoneTOT = 2.738555585555844;
    \effsysttrackingtwo = 2.735;
    \effsysttrackingtwort = 2.74;
    \effsysttrackingtwoTOT = 2.7382645771174303;
    \effsysttrackingthree = 2.736;
    \effsysttrackingthreert = 2.737;
    \effsysttrackingthreeTOT = 2.7369723710129037;
    \effsysttrackingfour = 2.74;
    \effsysttrackingfourrt = 2.739;
    \effsysttrackingfourTOT = 2.7390972675257794;
    \effsysttrackingfive = 2.744;
    \effsysttrackingfivert = 2.744;
    \effsysttrackingfiveTOT = 2.7438364977274303;
    \effsystpmodone = 0.049;
    \effsystpmodonert = 0.058;
    \effsystpmodoneTOT = 0.03691850709139006;
    \effsystpmodtwo = 0.027;
    \effsystpmodtwort = 0.019;
    \effsystpmodtwoTOT = 0.015193374042751267;
    \effsystpmodthree = 0.004;
    \effsystpmodthreert = 0.008;
    \effsystpmodthreeTOT = 0.005744014901373904;
    \effsystpmodfour = 0.019;
    \effsystpmodfourrt = 0.026;
    \effsystpmodfour = 0.01764469696047067;
    \effsystpmodfive = 0.05; 
    \effsystpmodfivert = 0.075; 
    \effsystpmodfiveTOT = 0.05097976201140251;
    \effsystpidone = 2.19;
    \effsystpidonert = 0.373;
    \effsystpidoneTOT = 1.0114618542823666;
    \effsystpidtwo = 1.784;
    \effsystpidtwort = 0.026;
    \effsystpidtwoTOT = 0.6804627021383836;
    \effsystpidthree = 2.207;
    \effsystpidthreert = 0.767;
    \effsystpidthreeTOT = 1.3016066578943317;
    \effsystpidfour  = 2.799;
    \effsystpidfourrt  = 0.035;
    \effsystpidfourTOT = 1.0485345501725256;
    \effsystpidfive = 2.918;
    \effsystpidfivert = 0.93;
    \effsystpidfiveTOT = 1.6705491983212053;
    \effsystdatamc = 0.4;
    \totalsystone = 100*1.1381980024/21.3687307111;
    \totalsysttwo = 100*1.1211742754/21.3949604176;
    \totalsystthree = 100*0.4054886142/6.286059659;
    \totalsystfour = 100*0.2686168135/1.7246551547;
    \totalsystfive = 100*0.0187220948/0.0899782231;
}

\usepackage{longtable} 

\begin{document}

\renewcommand{\thefootnote}{\fnsymbol{footnote}}
\setcounter{footnote}{1}


\begin{titlepage}
\pagenumbering{roman}

\vspace*{-1.5cm}
\centerline{\large EUROPEAN ORGANIZATION FOR NUCLEAR RESEARCH (CERN)}
\vspace*{1.5cm}
\noindent
\begin{tabular*}{\linewidth}{lc@{\extracolsep{\fill}}r@{\extracolsep{0pt}}}
\ifthenelse{\boolean{pdflatex}}
{\vspace*{-1.5cm}\mbox{\!\!\!\includegraphics[width=.14\textwidth]{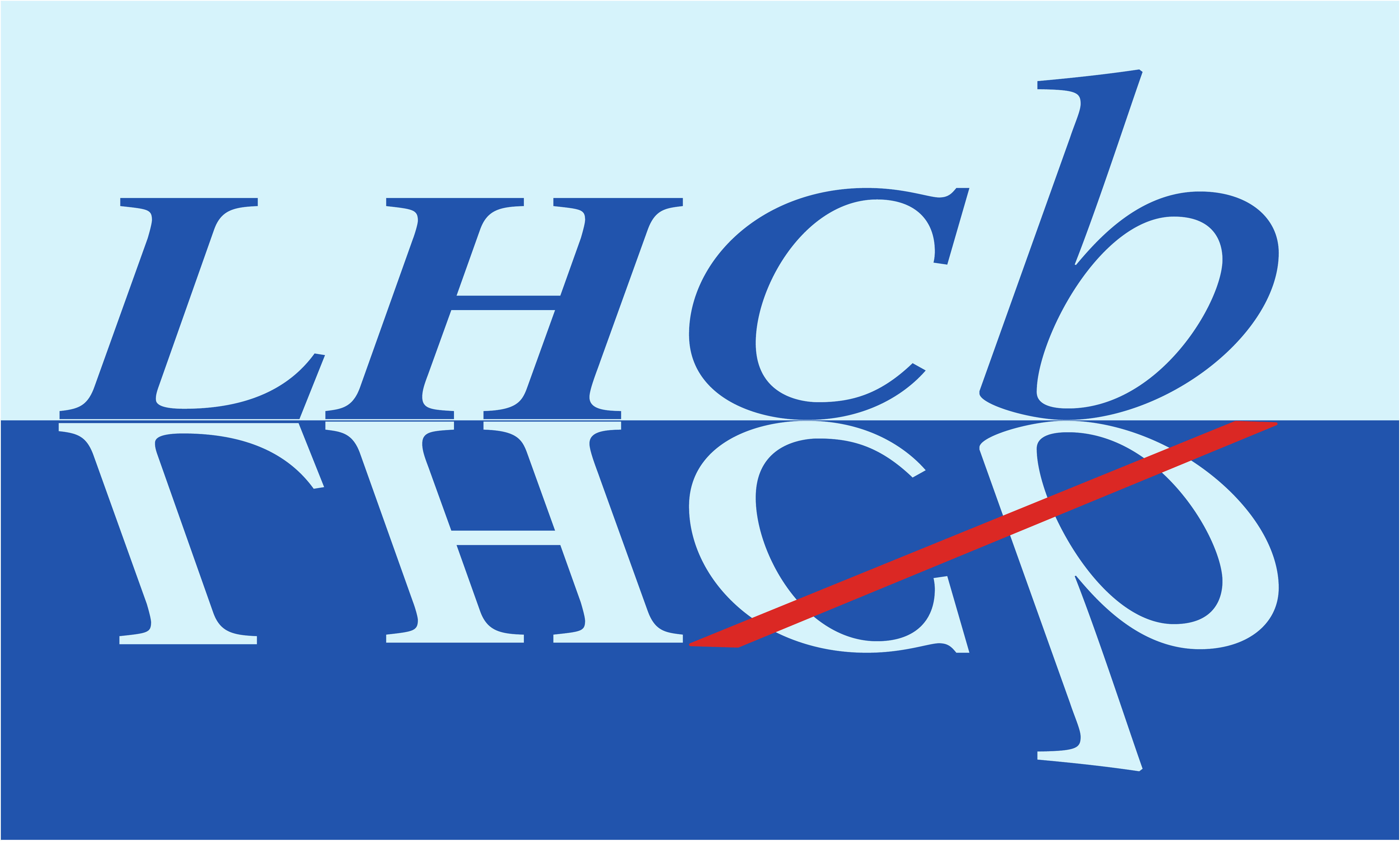}} & &}%
{\vspace*{-1.2cm}\mbox{\!\!\!\includegraphics[width=.12\textwidth]{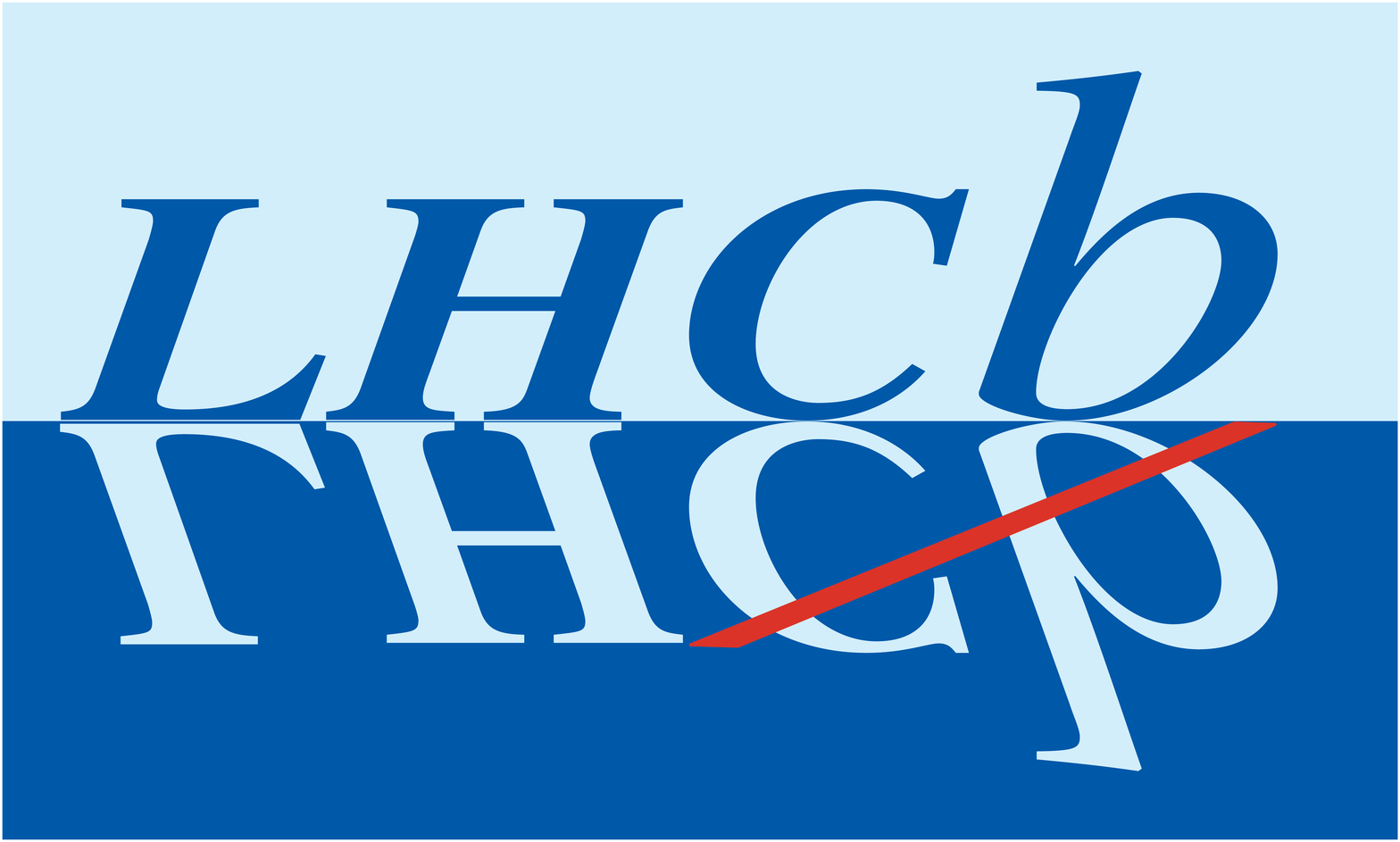}} & &}%
\\
 & & CERN-EP-2019-236 \\  
 & & LHCb-PAPER-2019-034 \\  
 & & March 25, 2020 \\ 
 & & \\
\end{tabular*}

\vspace*{4.0cm}

{\normalfont\bfseries\boldmath\huge
\begin{center}
  \papertitle 
\end{center}
}

\vspace*{2.0cm}

\begin{center}
\paperauthors\footnote{Authors are listed at the end of this paper.}
\end{center}

\vspace{\fill}

\begin{abstract}
  \noindent
The Cabibbo-suppressed semileptonic decay \btoppbarmunu is observed for the first time using a sample of $pp$ collisions corresponding to an integrated luminosity of 1.0, 2.0 and 1.7\invfb at centre-of-mass energies of 7, 8 and 13\tev, respectively. The differential branching fraction is measured as a function of the \ppbar invariant mass using the decay mode \bptojpsikp for normalisation.  The total branching fraction is measured to be
\begin{align*}
    \mathcal{B}(\btoppbarmunu) = (5.27 ^{+0.23}_{-0.24} \pm 0.21 \pm 0.15)\times 10^{-6},
\end{align*}
where the first uncertainty is statistical, the second systematic and the third is from the uncertainty on the branching fraction of the normalisation channel.
\end{abstract}

\vspace*{2.0cm}

\begin{center}
Published in JHEP 03 (2020) 146
\end{center}

\vspace{\fill}

{\footnotesize 
\centerline{\copyright~\papercopyright. \href{\paperlicenceurl}{\paperlicence}.}}
\vspace*{2mm}

\end{titlepage}


\newpage
\setcounter{page}{2}
\mbox{~}
%
%
%
%

\cleardoublepage


\renewcommand{\thefootnote}{\arabic{footnote}}
\setcounter{footnote}{0}



\pagestyle{plain} 
\setcounter{page}{1}
\pagenumbering{arabic}


%

\section{Introduction}
\label{sec:Introduction}

Studies of semileptonic $B$ meson decays have recently generated interest due to a number of anomalies in experimental results. Measurements of the observables $R(D)$ and $R(D^{\ast})$~\cite{LHCb-PAPER-2017-017,LHCb-PAPER-2015-025,Abdesselam:2019dgh,Hirose:2016wfn,Huschle:2015rga,Lees:2012xj} have shown hints of lepton non-universality with a combined significance of over \mbox{$3\,\sigma$~\cite{HFLAV19}}. To probe the flavour structure of possible new physics contributions to these decay modes, it is desirable to make analogous measurements for decays involving different quark-level processes, such as $\bquark\to\uquark$ transitions. To that end, the decay mode \bptoppbarlnu is promising experimentally, particularly when performing the measurement at a hadron collider. The requirement of a proton anti-proton pair in the final state should significantly reduce combinatorial background, which would otherwise be significant for final states with pions.

Semileptonic decays of $B$ mesons to a final state containing multiple baryons are as yet unobserved. A theoretical model of \bptoppbarlnu decays has been constructed with perturbative QCD~(pQCD)~\cite{Geng:2011tr}. This model is based on studies of several fully hadronic \btoyyx decays where $Y$ and $\overline{Y}'$ represent baryons and $X$ one or more mesons. By fitting the angular distributions and decay rates of the hadronic modes the authors of Refs.~\cite{Geng:2011tr, PhysRevD.74.094023, PhysRevD.78.054016} estimate the differential rate of \bptoppbarlnu decays. They also predict the total branching fraction of the \bptoppbarlnu decay to be \mbox{$(1.04\pm0.38)\times 10^{-4}$} for \mbox{$l=\mu,\,e$} leptons. This prediction motivated a search by the \belle collaboration for this channel that lead to evidence for the \mbox{$\Bp\to\ppbar\ep\neue$} decay mode with \mbox{$3.0\,\sigma$} significance~\cite{Tien:2013nga}. The branching fraction was measured to be $(8.2^{+3.7}_{-3.2}\pm 0.6)\times 10^{-6}$, one order of magnitude smaller than the prediction.

The measurements of the fully hadronic modes show features that merit further investigation. It is surprising that the branching fractions of decays of $B$ mesons to final states comprising only two baryons are suppressed compared to those of two baryons and one or more extra final state particles~\cite{PDG2018}. For example, the branching fraction of \mbox{$\Bz\to\ppbar$} is two orders of magnitude smaller than that of the similar \mbox{$\Bz\to\ppbar\pip\pim$} decay~\cite{LHCb-PAPER-2017-022,PDG2018}. Furthermore, the invariant-mass distributions of the baryon pair in \btoyyx decays show a characteristic accumulation at low values, called the threshold enhancement effect~\cite{Aubert:2007qea,Aubert:2006qx,Wang:2003iz,Aaij:2014tua}. Measurements of \bptoppbarlnu semileptonic decays provide the ideal environment for understanding the $\langle Y\bar{Y}'|(\bar{q}'b)_{V-A}|B\rangle$ matrix element that contributes to hadronic decay modes.

In this paper, the first observation of the decay \bptoppbarmunu is presented. As the dynamics of the transition are not known, the branching fraction is measured in bins of \ppbar invariant mass. These bins are then summed to obtain a measurement of the total branching fraction. The decay \mbox{$\Bp\to\jpsi\Kp$}, with \mbox{$\jpsi\to\mup\mun$}, is chosen as the normalisation mode as it is fully reconstructed and can pass similar selection requirements to the signal. The branching fraction within a bin $i$ is
\begin{align*}
    \mathcal{B}_{i}(\bptoppbarmunu) = \frac{N_{i}(\bptoppbarmunu)}{N(\bptojpsikp)}\times \frac{\epsilon(\bptojpsikp)}{\epsilon_{i}(\bptoppbarmunu)}\nonumber\\ 
    \times \mathcal{B}(\bptojpsikp) \times \mathcal{B}(\jpsi\to\mup\mun), 
\end{align*}
where $N_{i}(\bptoppbarmunu)$ is the yield of $\bptoppbarmunu\nonumber$ candidates in bin $i$, \mbox{$N(\bptojpsikp)$} is the total yield of $\bptojpsikp$ candidates and $\epsilon$ represents the product of the detector acceptance and the reconstruction and selection efficiencies of the two modes. The branching fractions of \mbox{$\bptojpsikp$} and \mbox{$\jpsi\to\mup\mun$} decays are taken from Ref.~\cite{PDG2018}.

The signal yields are extracted from fits to a variable called the corrected mass, which accounts for the unreconstructed neutrino in the signal decay. It is defined as~\cite{Abe:1997sb}
\begin{align}
    m_{\mathrm{corr}} = |p_{\perp}| + \sqrt{|p_{\perp}|^{2} + m_{\proton\antiproton\mu}^{2}},
\end{align}
where $|p_{\perp}|$ is defined as the magnitude of the reconstructed $\proton\antiproton\mup$ momentum transverse to the $B$ flight direction and $m_{\proton\antiproton\mu}^{2}$ is the square of the $\proton\antiproton\mup$ invariant mass.

This study uses the data collected with the LHCb detector in proton-proton collisions in 2011, 2012 and 2016. This corresponds to integrated luminosities of 1.0, 2.0 and 1.7\invfb at centre-of-mass energies of 7, 8 and 13\tev, respectively. The 2011 and 2012 data sets are treated together and collectively referred as the Run~1 data set. Charge conjugate processes are implied throughout this paper.

\section{Detector and simulation}
\label{sec:Detector}
The \lhcb detector~\cite{LHCb-DP-2008-001,LHCb-DP-2014-002} is a single-arm forward spectrometer covering the \mbox{pseudorapidity} range $2<\eta <5$, designed for the study of particles containing \bquark or \cquark quarks. The detector includes a high-precision tracking system consisting of a silicon-strip vertex detector surrounding the $pp$ interaction region~\cite{LHCb-DP-2014-001}, a large-area silicon-strip detector located upstream of a dipole magnet with a bending power of about $4{\mathrm{\,Tm}}$, and three stations of silicon-strip detectors and straw drift tubes~\cite{LHCb-DP-2013-003,LHCb-DP-2017-001} placed downstream of the magnet. The tracking system provides a measurement of the momentum, \ptot, of charged particles with a relative uncertainty that varies from 0.5\% at low momentum to 1.0\% at 200\gevc. The minimum distance of a track to a primary vertex (PV), the impact parameter (IP), is measured with a resolution of $(15+29/\pt)\mum$, where \pt is the component of the momentum transverse to the beam in\,\gevc. Different types of charged hadrons are distinguished using information from two ring-imaging Cherenkov (RICH) detectors~\cite{LHCb-DP-2012-003}. Photons, electrons and hadrons are identified by a calorimeter system consisting of scintillating-pad and preshower detectors, an electromagnetic and a hadronic calorimeter. Muons are identified by a system composed of alternating layers of iron and multiwire proportional chambers~\cite{LHCb-DP-2012-002}.

The online event selection is performed by a trigger~\cite{LHCb-DP-2012-004}, which consists of a hardware stage that performs some basic selection, followed by a software stage, which applies a full event reconstruction. At the first level, a track consistent with being a muon with significant \pt is required to be present in the event. Subsequently in the software stage, two tracks are required to form a secondary vertex with significant displacement from a $pp$ interaction vertex. A multivariate algorithm~\cite{BBDT} is used to identify vertices that are consistent with the decay of a \bquark hadron.

Simulation is used to determine the efficiency of the signal mode and estimate the shapes of the signal and several backgrounds modes in the fits to the \mcorr distribution. In the simulation, $pp$ collisions are generated using \pythia~\cite{Sjostrand:2006za,*Sjostrand:2007gs} with a specific \lhcb configuration~\cite{LHCb-PROC-2010-056}.  Decays of unstable particles are described by \evtgen~\cite{Lange:2001uf}, in which final-state radiation is generated using \photos~\cite{Golonka:2005pn}. The interaction of the generated particles with the detector, and its response, are implemented using the \geant toolkit~\cite{Allison:2006ve, *Agostinelli:2002hh}, as described in Ref.~\cite{LHCb-PROC-2011-006}. The generated $B$ meson \ptot and \pt spectra are corrected to match the data distributions. A boosted decision tree (BDT) weighter~\cite{Rogozhnikov:2016bdp} is trained on samples of \bptojpsikp data and simulation, independent of those used for the normalisation of the branching fraction. This is then used to correct all the simulation samples used in the analysis.

\section{Selection}
\label{sec:Selection}

Signal candidates are constructed from three charged tracks which are required to be of good quality and well separated from any PV. The tracks must also have particle identification consistent with their particle hypothesis. The requirement for positive proton identification enforces a minimum value of \ptot of 18\gevc such that the protons are above the threshold for radiating in the RICH. Similarly, the muons must have \ptot above 3\gevc to propagate through the muon stations. All the tracks must have \pt larger than 1.5\gevc and form a good-quality vertex significantly displaced from the PV with which the candidate is associated. The signal muon must have fired the hardware trigger and the reconstructed \Bp candidate formed by the three tracks must be consistent with the object that fired the software trigger. Potential decays of $\eta_{c}$, $\jpsi$ and $\psi(2S)$ mesons to \ppbar are removed with vetoes in the $\ppbar$ invariant mass of $\pm50\mevcc$ around their respective known masses~\cite{PDG2018}.

The selection of the \bptojpsikp normalisation mode is aligned with that of the signal to reduce systematic uncertainties. The selection criteria for the signal protons are applied to the kaon and the muon of opposite sign (\Kp\mun), with the exception of the particle identification criteria. The selection of the other muon is the same as that of the muon in the signal decay.

Further selection is used to reduce several sources of backgrounds relative to the signal. In total there are five variables to which selection is applied, with the chosen criteria on each optimised together. These variables, and the backgrounds targeted by them are described in the following paragraphs.

The largest background contribution comes from a mixture of partially reconstructed decays producing two protons and a muon in the final state. It is expected that the largest among these originates in $\bquark\to\cquark$ quark transitions. The most pernicious is $B\to\Lcbar\proton\mup\neum X$ decays, where $X$ represents any number of charged or neutral pions (including none) and the \Lcbar baryon decays to a final state including one proton. The other major background arises from $B\to\ppbar \Dbar X$ decays, where the $\Dbar$ meson may be of any variety (\Dzb, \Dm, \Dstarm, \etc) and ultimately decays to a final state with a muon. The contribution of $B\to\proton\Lcbar X$ decays with the \Lcbar baryon decaying semileptonically is comparatively small, as the semileptonic branching fraction is dominated by $\Lcbar\to\Lbar l^{-}\bar{\nu}_{l}$ decays. The \Lz baryon flies a sufficient distance within the detector before decaying such that the resulting proton is not associated with the $B$ decay vertex. Another source of partially reconstructed background is formed of $B\to\ppbar\mup\neum X$ decays, where $X$ denotes one or more charged or neutral pions. These decays may proceed with intermediate $N^{\ast}$ or $\Deltares$ resonances and could naively be expected to have similar branching fractions to the signal.

\begin{figure}[t]
    \centering
    \begin{overpic}[width = 0.49\textwidth]{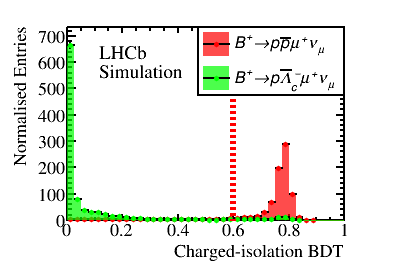}
    \put(30,45){(a)}
    \end{overpic}
    \begin{overpic}[width = 0.49\textwidth]{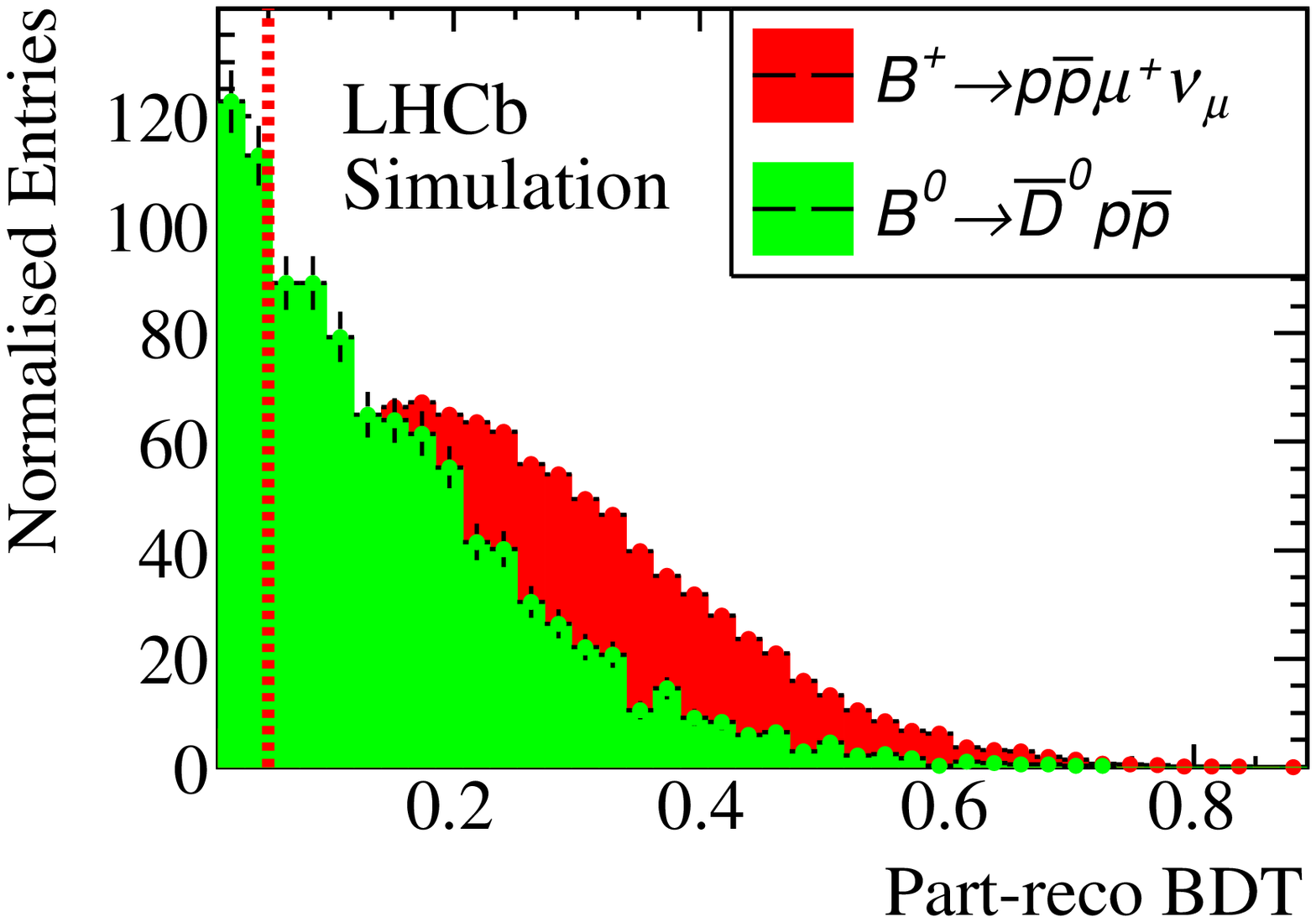}
    \put(30,45){(b)}
    \end{overpic}
    \caption{Result of training (a)~the charged-isolation BDT and (b)~the part-reco BDT. The chosen selection on the classifier outputs are indicated by the dashed red line. For some candidates there are no additional tracks near the $B$-decay vertex; these candidates are accepted and do not appear in the charged-isolation BDT output.
    The background samples shown here have the $\Lcbar$ and $\Dzb$ hadrons decaying via $\Lcbar\to\antiproton\Kp\pim$ and $\Dzb\to\mup X$. The part-reco BDT is trained on a mixture of background modes with only one shown here for illustration.}
    \label{sec:Selection:fig:BDTs}
\end{figure}

If any of these partially reconstructed decay modes produces charged tracks in addition to the \proton\antiproton\mup candidate, it can be efficiently suppressed with an isolation technique. Once a signal candidate has been constructed, the other tracks in the event close to the $B$ decay vertex are examined. A BDT is used to identify those nearby tracks that can be associated with the signal candidate decay vertex. If the candidate is truly signal, there should be few other tracks that can be associated with it and the BDT should classify them with a low score. On the other hand, the extra track(s) from a partially reconstructed decay returns a high score if such tracks are found. The isolation algorithm returns the BDT output for the four tracks most likely to have come from the $B$ vertex. These four numbers are themselves combined into a single BDT classifier, known as the charged-isolation BDT. This BDT is trained on simulation to discriminate signal from $\Bp\to\Lcbar\proton\mup\neum$ decays, which is expected to be the largest mode with extra charged tracks. The efficacy of this BDT in reducing such background is shown in Fig.~\ref{sec:Selection:fig:BDTs}(a). The indicated requirement on the charged BDT score rejects 80\% of the major background decay $B\to\Lcbar\proton\mup\neum X$ (with all possible decay modes of the \Lcbar considered), whilst retaining 93\% of the signal.

For those partially reconstructed final states with only additional neutral particles, further suppression is achieved by considering the kinematics of the decays. An additional BDT, the so called part-reco BDT, considers 11 variables: the impact parameter significance of the three final-state tracks, the \ppbar pair and the \Bp candidate with respect to the PV; the impact parameters of the tracks with respect to the fitted \Bp decay vertex; the $\chi^{2}$ of the \Bp vertex fit; the angle between the \Bp candidate momentum and displacement vectors; and the difference between the \proton and \antiproton momenta. The part-reco BDT is trained on simulation in order to discriminate signal from a mixture of all the considered background modes. The result of this training is shown in Fig.~\ref{sec:Selection:fig:BDTs}(b). The selection on the part-reco BDT output removes 18\% of the decays $B\to\proton\antiproton \Dbar$ and keeps 98\% of the signal.

An additional background arises from particles that are misidentified as protons (misID). The particle identification requirements on the proton tracks are therefore further tightened. Background due to hadrons being misidentified as muons is considered and reduced to a negligible amount with a loose particle identification requirement. A background occurs due to the combination of two tracks from the decay of a heavy hadron with a track from elsewhere in the event. This is referred to as combinatorial background. This component is expected to have a small contribution due to the tight vertex requirements on the $\proton\antiproton\mup$ candidate and the requirement for positively identifying two protons. Therefore no additional selection is employed specifically to reduce it.

In addition to the two BDTs and proton identification criteria, one further quantity is considered. The uncertainty on the corrected mass of the candidate may be used to improve the separation between signal and background~\cite{LHCb-PAPER-2015-013}. It is calculated from the estimated uncertainties on the positions of the \Bp primary and secondary vertices, and the momenta of the tracks. Selecting lower values of the corrected-mass uncertainty produces a sharper peak for the signal mode in the corrected mass distribution, which will aid the discrimination of the signal from background in the fit to determine the yield.

In total the selection uses five quantities (two BDTs, the proton PID, the muon PID and the corrected-mass uncertainty). In order to ascertain the optimum selection, a five dimensional grid search is performed using pseudoexperiments. Data sets are generated from the simulation samples with the expected proportions of each background. The expected signal amount is taken from the central value of the $\Bp\to\proton\antiproton\ep\neue$ branching fraction reported by the \belle collaboration~\cite{Tien:2013nga}. For the backgrounds, the current averages for the branching fractions are used if they have been measured. For those backgrounds that have not been measured, their branching fractions are estimated relative to that expected for the signal, accounting for different CKM matrix elements and the available phase space. For each point in the grid, the selection is applied to the simulation to estimate the efficiency for each component. The efficiency of the PID requirements on the simulation is estimated with a method based on data control samples~\cite{LHCb-PUB-2016-021}. For each data set the \mcorr variable is simulated and the expected relative uncertainty on the signal yield is found by a fit to the simulated pseudodata. These fits are not binned in \mppbar but consider the entire sample. The selection that produces the smallest relative uncertainty on the signal yield is chosen.

\section{Signal and normalisation yields}
\label{sec:Fit}

\begin{figure}[t]
    \centering
    \includegraphics[width = 0.49\textwidth]{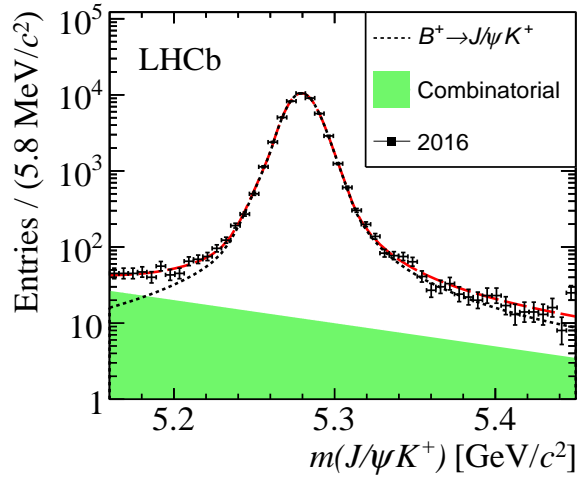}
    \caption{Distribution of $m(\jpsi\Kp)$ with the fit result shown for the 2016 data set.}
    \label{sec:Fit:fig:normFit}
\end{figure}

The yields of the signal and normalisation modes are ascertained with unbinned extended maximum-likelihood fits. In the case of the normalisation mode, the invariant mass distribution of the $\jpsi\Kp$ candidates is fitted. The 2011, 2012 and 2016 data sets are fitted separately and then the yields combined. The fit to the 2016 data set is shown in Fig.~\ref{sec:Fit:fig:normFit}.

For the signal mode, the corrected mass is fitted. The distribution of this variable peaks at the \Bp mass for candidates where one massless particle is not reconstructed. On the other hand, candidates from partially reconstructed decays that are missing one or more massive particles in addition to the neutrino have wide distributions concentrated at lower corrected mass values. The Run~1 and 2016 data are combined and fitted together to improve the fit stability.

The shapes for the signal component and contributions from partially reconstructed decays are determined using simulation. The shape of the signal probability density function (PDF) is parametrised by the sum of four bifurcated Gaussian functions with a common mean. The parameters of the Gaussian functions as well as their relative fractions are all fixed in the fit. All of the background PDFs are accounted for with kernel density estimation~\cite{Cranmer:2000du}.

The shape of the proton misID background comes from a separate independent data sample in which the particle identification requirements on one of the protons have been removed. In this sample the true number of each hadron species can be unfolded and so the probability of a hadron being misidentified as a proton can be estimated. These probabilities are used to weight this sample to estimate both the template shape for the fit component and the yield of misID events.

A background component due to random combinations of protons and muons, referred to as the combinatorial background, is included in the fit. A sample of data for which the \Bp decay vertex quality selection has been reversed is used to estimate the shape of this background. 

The yields of the signal, proton misID, combinatorial and partially reconstructed decays are determined by the fit, as are the relative fractions of each partially reconstructed mode. All of the fit parameters are free to vary with the exception of the misID yield which is constrained.

\begin{figure}[htbp]
    \centering
    \includegraphics[width = 0.49\textwidth]{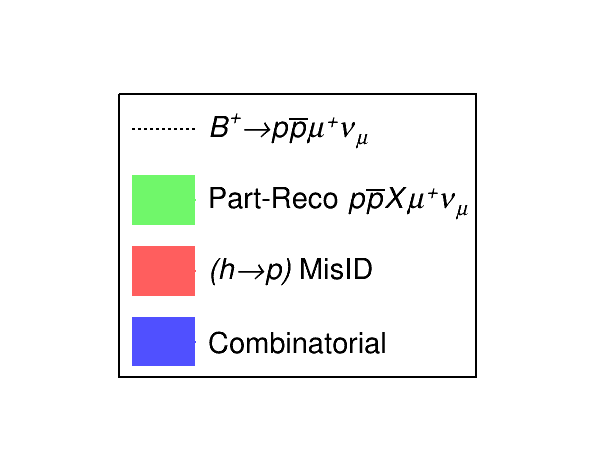}
    \includegraphics[width = 0.49\textwidth]{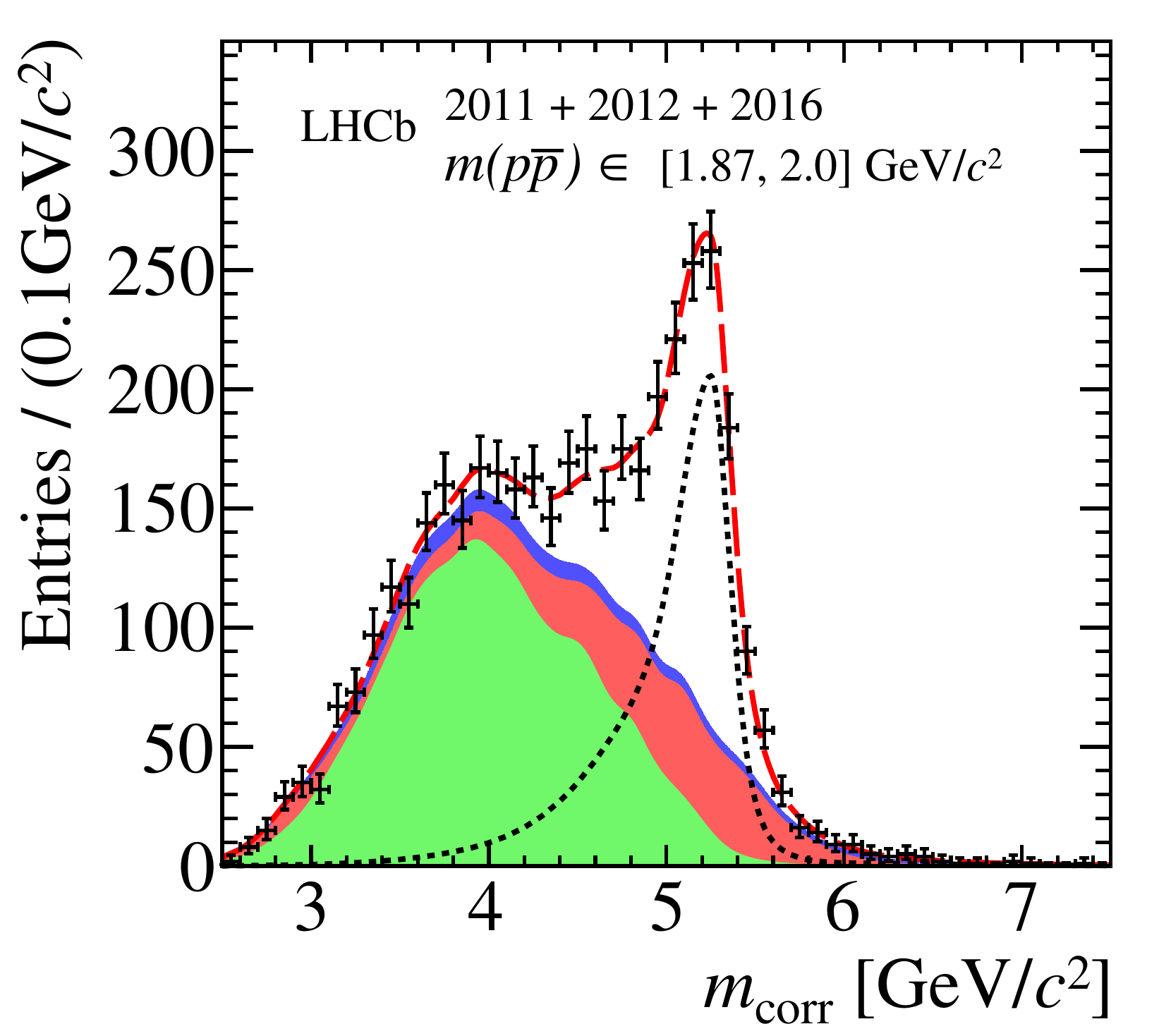}\\
    \includegraphics[width = 0.49\textwidth]{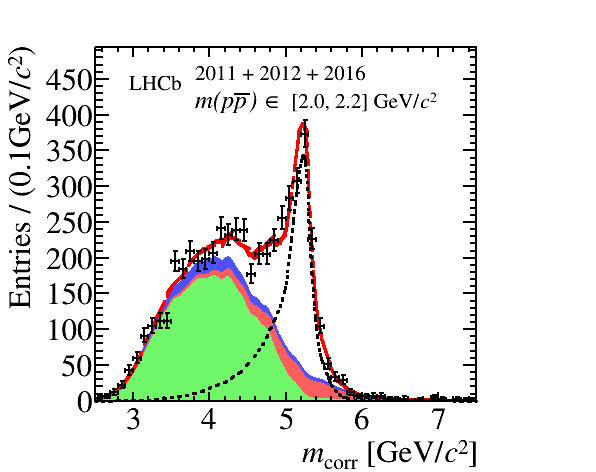}
    \includegraphics[width = 0.49\textwidth]{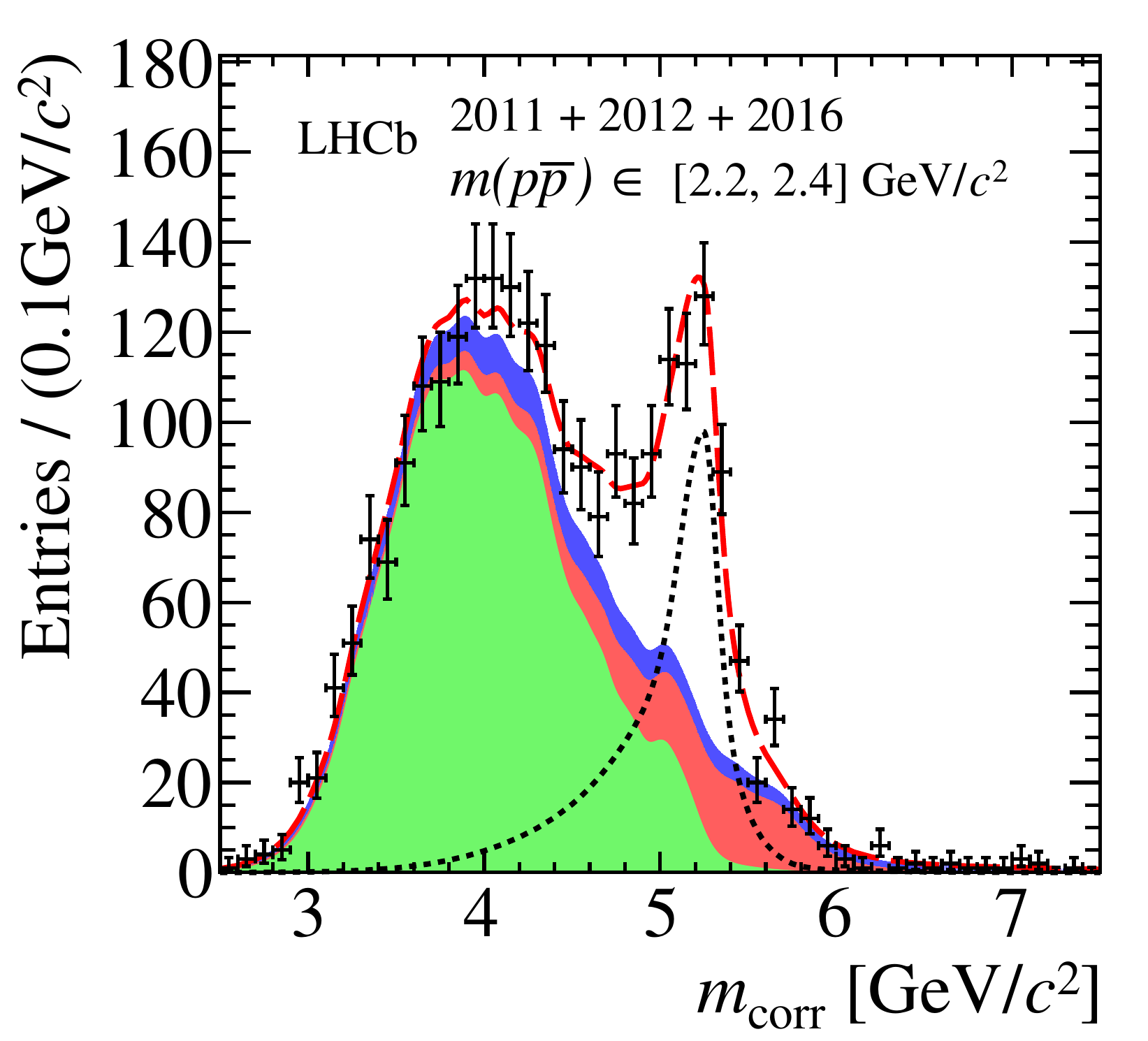}\\
    \includegraphics[width = 0.49\textwidth]{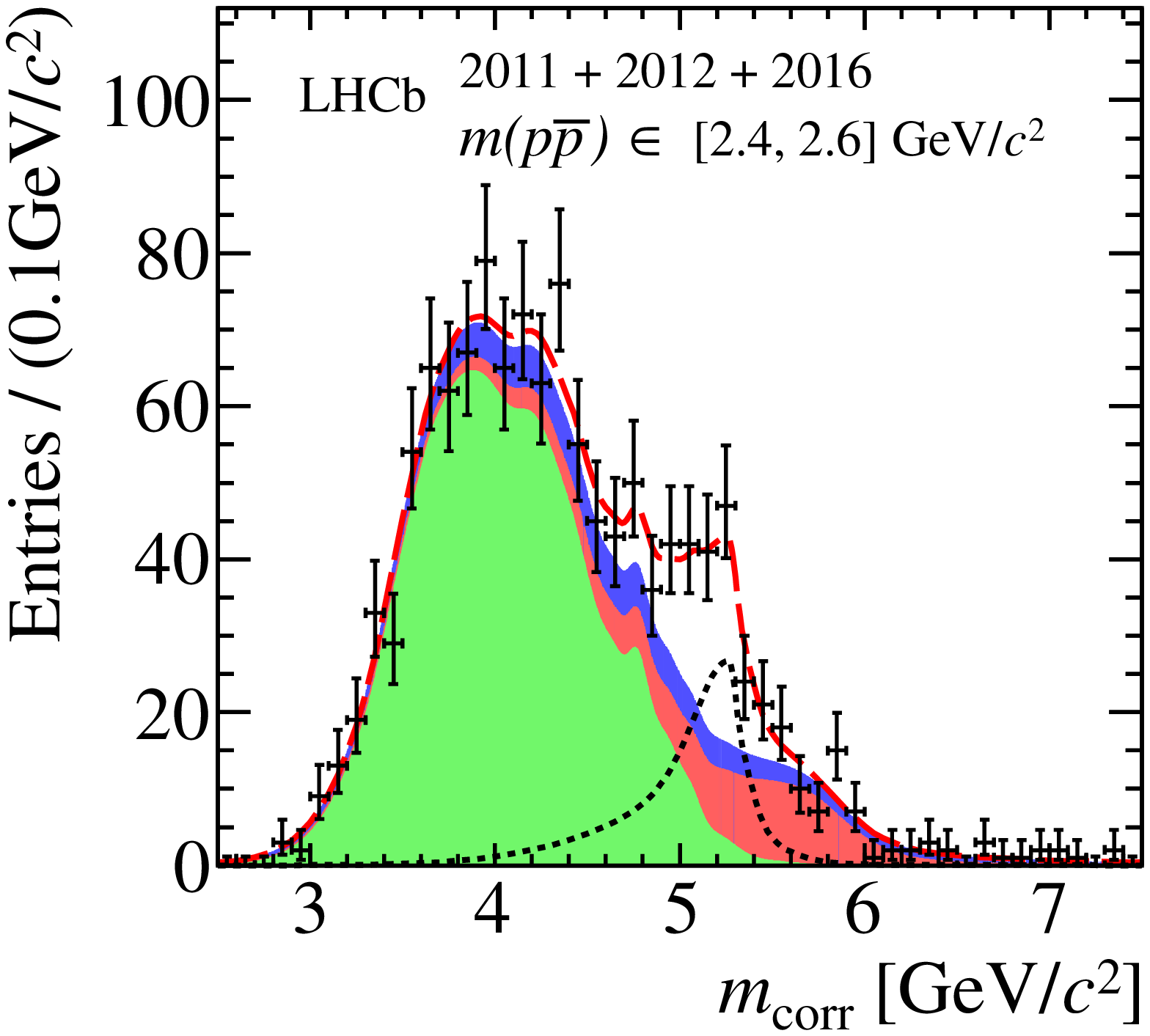}
    \includegraphics[width = 0.49\textwidth]{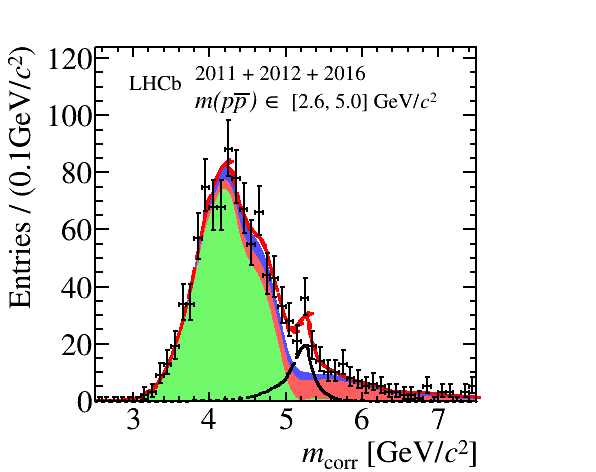}
    \caption{Distributions of \mcorr in each \mppbar bin with the fit results shown.}
    \label{sec:Fit:fig:mcorr}
\end{figure}

The fit in each \mppbar bin is performed independently. The \mcorr distributions in each bin, and the resulting fits are shown in Fig.~\ref{sec:Fit:fig:mcorr}. In each bin the fits are validated using pseudoexperiments. An ensemble of $10^{5}$ data sets is generated and fitted with the component yields taken from the fits to data. Some small biases on the signal yield are found and these are considered as a source of systematic uncertainty.

\section{Efficiency}
\label{sec:Efficiency}

The efficiencies for the signal and normalisation modes to be reconstructed and selected are both assessed with simulation. Corrections are applied to account for known differences between data and simulation in the track-reconstruction efficiency~\cite{track_eff} and the efficiency of the hardware trigger~\cite{LHCb-PUB-2014-039}. The efficiency of the particle identification requirements on each track is evaluated with data~\cite{LHCb-PUB-2016-021} and applied to the simulation.

The binning in \mppbar reduces the dependence on the model of the \Bp decay when calculating the efficiency of the signal mode. However, as there are selection requirements on kinematic quantities of the candidates, there is still some residual dependence on the dynamics of the decay. The simulation is therefore weighted to represent the pQCD model of Ref.~\cite{Geng:2011tr} as the current best estimate of how the decay proceeds. This weighting corrects the distribution of the invariant mass of the $\mup\neum$ system. The variation of the parameters of this model is considered as a source of systematic uncertainty.

The ratio of selection efficiencies between the signal and normalisation modes in each bin of \mppbar is shown in Table~\ref{sec:Efficiency:tab:effs}. These efficiencies are presented separately for the Run~1 and 2016 samples. They are combined to form an overall efficiency ratio, accounting for the difference in sample sizes between Run~1 and 2016. This combination is calculated using the measured \Bp production cross-sections~\cite{LHCb-PAPER-2016-031} and integrated luminosities of each data set.

\begin{table}[t]
     \centering
          \caption{Relative efficiencies for Run~1 and 2016 and the weighted combination of both.}
     \begin{tabular}{ c c c c }\toprule
        \multirow{2}{*}{$m(\ppbar)_i$ [\gev/c$^{2}$]} & \multicolumn{3}{c}{ $\epsilon(B\to \ppbar \mu \nu)_{i}/\epsilon(B\to \jpsi K)$} \\  
 & Run~1 & 2016 & Run~1 \& 2016 \\
\midrule
Bin~1: $1.87-2.0$ &  0.37 $\pm$ 0.02 & 0.57 $\pm$ 0.03 & 0.48 $\pm$ 0.02 \\ 
Bin~2: \phantom{0}$2.0-2.2$ &  0.37 $\pm$ 0.02 & 0.51 $\pm$ 0.03 & 0.45 $\pm$ 0.02 \\
Bin~3: \phantom{0}$2.2-2.4$ &  0.36 $\pm$ 0.02 & 0.50 $\pm$ 0.03 & 0.44 $\pm$ 0.02 \\
Bin~4: \phantom{0}$2.4-2.6$ &  0.36 $\pm$ 0.02 & 0.52 $\pm$ 0.03 & 0.45 $\pm$ 0.02 \\
Bin~5: \phantom{0}$2.6-5.0$ &  0.35 $\pm$ 0.02 & 0.49 $\pm$ 0.02 & 0.43 $\pm$ 0.02 \\
\bottomrule
 \end{tabular}
     \label{sec:Efficiency:tab:effs} 
 \end{table}

\section{Systematic uncertainties}
\label{sec:systematics}

The systematic uncertainties can be split into two categories: those that affect the calculation of the ratio of efficiencies of the signal and normalisation modes and those that may change the determination of the signal yield in the fit. For the former, each of the corrections to the simulation contributes a source of uncertainty both from the limited sizes of the samples used to derive the corrections and from the method of deriving them. The method of correcting the \ptot and \pt distributions of the \Bp mesons in the simulation may give rise to a systematic uncertainty. The parameters of the BDT weighter used to correct these distributions are altered and the relative efficiencies recalculated, with the difference to the nominal relative efficiency being the assigned uncertainty. An additional uncertainty due to any residual differences between data and simulation is determined using the $\Bp\to\jpsi\Kp$ decay mode. The difference in efficiency due to the selection on the two BDTs and corrected-mass uncertainty is compared between data and simulation.

To account for the uncertainty in the correction of the simulation for the reconstruction efficiency of each track, the applied weights are varied within their uncertainties and the relative efficiencies recalculated. Similarly, an uncertainty is assessed for the particle-identification weights applied to each track. The uncertainty due to the limited simulation sample sizes used to calculate the efficiencies is also included.

A further uncertainty is due to the physics model that the simulation is weighted to represent. The model affects the kinematic distributions of the final state tracks which feeds into the efficiency calculation as these distributions are biased by the selection requirements. Since the model is unproven a conservative uncertainty is taken. New sets of weights for the simulation are created that sample extreme variations of the model parameters ($\pm5\,\sigma$), and for each variation the efficiency is recalculated. Despite this extreme test, the systematic uncertainty due to the physics model is not dominant, which reflects the flat selection efficiency over the kinematic ranges in which the final-state particles lie within each bin of \mppbar. Finally, the uncertainties on the \Bp production cross-section~\cite{LHCb-PAPER-2016-031} and integrated luminosities of the data samples are combined to give the systematic uncertainty on the averaging of the efficiencies when combining Run~1 and 2016.

In the corrected-mass fit, uncertainties arise from potential variations in the shapes of the components. This variation is assessed with pseudoexperiments. Data sets are generated with the nominal fit model and then fitted with both the nominal model and an alternative. The width of the distribution of differences between the nominal and alternative fits is taken as the uncertainty. For those components that rely on kernel density estimators, a systematic uncertainty is assessed for the choice of smoothing parameter by varying it. The uncertainty due to the choice of model for the signal shape is found by replacing the nominal PDF with one constructed with kernel density estimators. The uncertainty due to the limited sizes of the simulation samples is determined by generating new simulation from the nominal fit PDFs with the same sample sizes and making alternative PDFs with those samples. Similarly, an estimate of the uncertainty on the shape of the proton misID background component is assessed. For the shape of the combinatorial background component, an alternative data sample is trialled which requires the two protons to be of the same charge. Finally, the small biases in the fit noted in Sec.~\ref{sec:Fit} are included.

A summary of the systematic uncertainties is presented in Table~\ref{sec:systematics:tab:summary}. They are given as relative uncertainties on the branching fraction with the combination accounting for the correlation of the uncertainties between the two data sets. The correlations of the total uncertainties between the bins are shown in Table~\ref{sec:appendix:table:corr} and the covariance matrix is presented in Table~\ref{sec:appendix:table:covar}, in the appendix.

\begin{table}[t]
    \centering
        \caption{Summary of the systematic uncertainties on the differential branching fractions. The contributions pertaining to the efficiency estimate are first, those for the yield extraction are below. The particle identification and tracking efficiency uncertainties are assumed to be 100\% correlated between Run~1 and 2016. The total correlations of the uncertainties between the bins are shown in Table~\ref{sec:appendix:table:corr}.}
    \begin{tabular}{c c c c c c}
    \toprule
    \multirow{2}{*}{Source} & \multicolumn{5}{c}{Relative uncertainties on $\mathcal{B}$ [\%]}\\
    & Bin 1 & Bin 2 & Bin 3 & Bin 4 & Bin 5\\
        \midrule
         Kinematic weighting &
       0.7 &
       0.6 &
       \phantom{0}0.4 &
       \phantom{0}0.5 &
       \phantom{0}0.4\\
         Data-simulation agreement &
          0.4 &
          0.4 &
          \phantom{0}0.4 &
          \phantom{0}0.4 &
          \phantom{0}0.4\\
          Tracking efficiency & 
         2.7 &
         2.7 &
         \phantom{0}2.7 &
         \phantom{0}2.7 &
         \phantom{0}2.7\\
         Particle identification & 
         1.0 &
         0.7 &
         \phantom{0}1.3 &
         \phantom{0}1.0 &
         \phantom{0}1.7\\
         Simulation sample size & 
         3.6 &
         3.2 &
         \phantom{0}3.2 &
         \phantom{0}3.1 &
         \phantom{0}3.0\\
         Physics model & 
         0.3 &
         0.6 &
         \phantom{0}0.6 &
         \phantom{0}0.4 &
         \phantom{0}0.3\\
         Run 1 and 2016 combination &
         2.1 &
         1.6 &
         \phantom{0}1.7 &
         \phantom{0}1.7 &
         \phantom{0}1.6\\
         \midrule
         Kernel smoothing &
         0.0 &
         1.1 &
         \phantom{0}2.7 &
         \phantom{0}7.9 &
         \phantom{0}3.5\\
         Signal model &
         0.6 &
         2.0 &
         \phantom{0}3.0 &
         \phantom{0}4.8 &
         \phantom{0}9.9\\
         Simulation sample size &
         0.3 &
         0.0 &
         \phantom{0}0.3 &
         \phantom{0}2.4 &
         \phantom{0}5.2\\
         misID model &
         0.9 &
         0.1 &
         \phantom{0}0.6 &
         \phantom{0}5.2 &
         13.5\\
         Combinatorial model &
         0.9 &
         1.2 &
         \phantom{0}1.2 &
         \phantom{0}8.5 &
         \phantom{0}4.7\\
         Fit bias &
         0.2 &
         0.1 &
         \phantom{0}0.9 &
         \phantom{0}2.5 &
         \phantom{0}7.8\\
         \midrule
         Total systematic uncertainty &
         5.3 &
         5.2 &
         \phantom{0}6.5 &
         15.6 &
         20.8\\
         \midrule
         Total statistical uncertainty &
         9.1 &
         5.5 &
         12.5 &
         25.3 &
         29.8\\
         \bottomrule
    \end{tabular}
    \label{sec:systematics:tab:summary}
\end{table}

\section{Results}
\label{sec:Results}

\begin{figure}[tbp]
    \centering
    \includegraphics[width=0.7\linewidth]{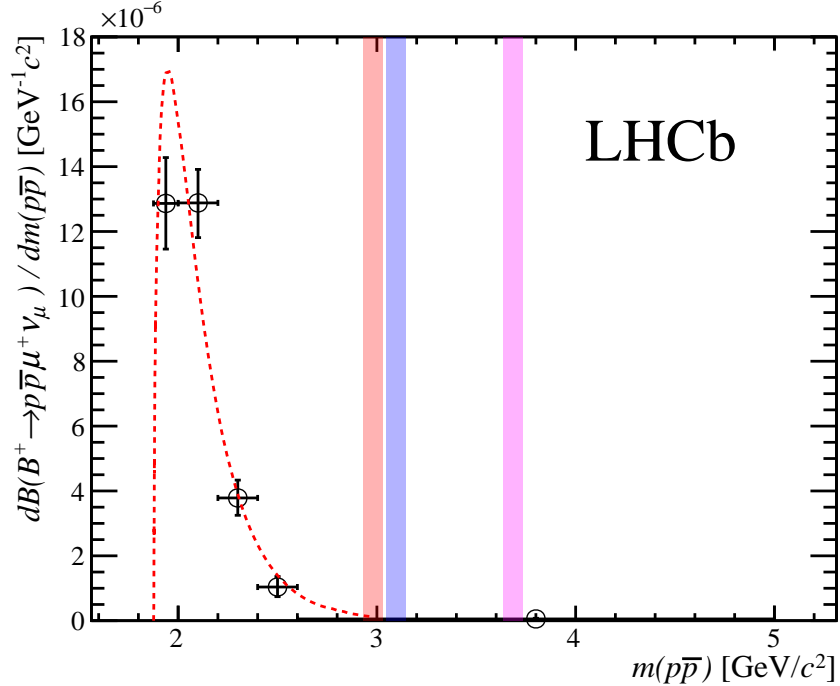}
\caption{Differential branching fraction as a function of the \ppbar invariant mass. The \mbox{$\etac\to\proton\antiproton$}, \mbox{$\jpsi\to\proton\antiproton$} and \mbox{$\psitwos\to\proton\antiproton$} vetoes are indicated by the (left) red, (middle) blue and (right) pink bands, respectively. The red dashed line represents the prediction of the pQCD model normalised to the observed branching fraction~\cite{Geng:2011tr}.}
\label{sec:Results:fig:dbf}
\end{figure}

The fitted yields for the signal mode are presented in Table~\ref{sec:Results:tab:sigyields}. The extracted yields of the normalisation channel are $14\,930\pm 260$ for 2011, $31\,380 \pm 190$ for 2012 and \mbox{$49\,270 \pm 250$} for 2016. Combining these with the efficiency ratios from Sec.~\ref{sec:Efficiency}, the differential branching fraction in each \mppbar bin is calculated. The results are presented in Table~\ref{sec:Results:tab:sigyields}. The relative differential branching fractions are summed over the bins, with the correlation of the systematic uncertainties between the bins accounted for, to give the total relative branching fraction of
\begin{align*}
    \frac{\mathcal{B}(\bptoppbarmunu)}{\mathcal{B}(\bptojpsikp)\times\mathcal{B}(\jpsi\to\mup\mun)} = (8.75\pm0.39\pm 0.35)\times 10^{-2},
\end{align*}
where the first uncertainty is statistical and the second systematic. Multiplying this by the current average of the normalisation branching fraction~\cite{PDG2018} leads to
\begin{align*}
    \mathcal{B}(\bptoppbarmunu) = (5.27^{+0.23}_{-0.24}\pm 0.21 \pm 0.15)\times 10^{-6},
\end{align*}
where the third uncertainty is from the normalisation branching fraction. Finally, the absolute differential branching fraction as a function of \mppbar is shown in Fig.~\ref{sec:Results:fig:dbf}, where the indicated uncertainties include statistical, systematic and normalisation uncertainty contributions. As expected from the theory model and the analogous hadronic decays, the differential distribution peaks at a very low value and falls off sharply. The measured total branching fraction agrees with the previous measurement from the \belle collaboration and represents the first observation of the \btoppbarmunu decay mode.

\begin{table}[tbp]
\renewcommand*{\arraystretch}{1.4}
     \centering
          \caption{Number of observed \btoppbarmunu candidates and differential branching fraction in each bin of \mppbar. The uncertainties on the signal yields are statistical only. For the differential branching fractions the first uncertainties are statistical, the second systematic and the third from the uncertainties on the branching fractions of the normalisation channel.}
     \begin{tabular}{c c c}
          \toprule
          \mppbar [\gev/c$^{2}$] & Signal Yield & $d\mathcal{B}(\bptoppbarmunu)/d{\mppbar}$ [$\times10^{-6}\gev^{-1}$c$^2$]\\
\midrule
Bin~1: $1.87-2.0$ & $1210 \pm 110$ & \phantom{0}$12.9 \pm 1.2 \pm 0.7 \pm 0.4$\\
Bin~2: \phantom{0}$2.0-2.2$ & $ 1830 \pm 110$ & \phantom{0}$12.9 \pm 0.7 \pm 0.7 \pm 0.4$\\
Bin~3: \phantom{0}$2.2-2.4$ & $ 530 \pm 70 $ & \phantom{00}$3.8 \pm 0.5\pm 0.2 \pm 0.1$\\
Bin~4: \phantom{0}$2.4-2.6$ & $150 \pm 40$ & \phantom{0}$1.04 \pm 0.30 \pm 0.16  \pm 0.03$\\
Bin~5: \phantom{0}$2.6-5.0$ & \phantom{0}$88 \pm 26$ & $0.054 \pm 0.016 \pm 0.011 \pm 0.002$\\
\bottomrule
 \end{tabular}
     \label{sec:Results:tab:sigyields} 
\end{table}

\section*{Acknowledgements}
%
%
\noindent We express our gratitude to our colleagues in the CERN
accelerator departments for the excellent performance of the LHC. We
thank the technical and administrative staff at the LHCb
institutes.
We acknowledge support from CERN and from the national agencies:
CAPES, CNPq, FAPERJ and FINEP (Brazil); 
MOST and NSFC (China); 
CNRS/IN2P3 (France); 
BMBF, DFG and MPG (Germany); 
INFN (Italy); 
NWO (Netherlands); 
MNiSW and NCN (Poland); 
MEN/IFA (Romania); 
MSHE (Russia); 
MinECo (Spain); 
SNSF and SER (Switzerland); 
NASU (Ukraine); 
STFC (United Kingdom); 
DOE NP and NSF (USA).
We acknowledge the computing resources that are provided by CERN, IN2P3
(France), KIT and DESY (Germany), INFN (Italy), SURF (Netherlands),
PIC (Spain), GridPP (United Kingdom), RRCKI and Yandex
LLC (Russia), CSCS (Switzerland), IFIN-HH (Romania), CBPF (Brazil),
PL-GRID (Poland) and OSC (USA).
We are indebted to the communities behind the multiple open-source
software packages on which we depend.
Individual groups or members have received support from
AvH Foundation (Germany);
EPLANET, Marie Sk\l{}odowska-Curie Actions and ERC (European Union);
ANR, Labex P2IO and OCEVU, and R\'{e}gion Auvergne-Rh\^{o}ne-Alpes (France);
Key Research Program of Frontier Sciences of CAS, CAS PIFI, and the Thousand Talents Program (China);
RFBR, RSF and Yandex LLC (Russia);
GVA, XuntaGal and GENCAT (Spain);
the Royal Society
and the Leverhulme Trust (United Kingdom).

\pagebreak

\section*{Appendix}
\label{sec:appendix}
\appendix
\section{Correlation and covariance matrices}
\begin{table}[htbp]
\renewcommand*{\arraystretch}{1.4}
     \centering
     \caption{Correlations in the uncertainties between bins of \mppbar.}
     \begin{tabular}{ c | c c c c c }
     & Bin 1 & Bin 2 & Bin 3 & Bin 4 & Bin 5\\
\mppbar [\gevcc] & $1.87-2.0$ & $2.0-2.2$ & $2.2-2.4$ & $2.4-2.6$ & $2.6-5.0$\\
 \hline
 $1.87-2.0$ & 1.00  & 0.19  & 0.11  &  0.05 &  0.04 \\
 \phantom{0}$2.0-2.2$ & - &  1.00  &  0.15 &  0.07 & 0.06 \\
 \phantom{0}$2.2-2.4$ & - & - & 1.00 & 0.04 & 0.04 \\
 \phantom{0}$2.4-2.6$ & - & - & - & 1.00 & 0.02 \\
 \phantom{0}$2.6-5.0$ & - & -  & - &  -  & 1.00 \\
 \end{tabular}
\label{sec:appendix:table:corr}
\end{table}

\begin{table}[htbp]
\renewcommand*{\arraystretch}{1.4}
     \centering
     \caption{Covariance matrix for bins of \mppbar.}
     \begin{tabular}{ c | c c c c c }
     & Bin 1 & Bin 2 & Bin 3 & Bin 4 & Bin 5\\
\mppbar [\gevcc] & $1.87-2.0$ & $2.0-2.2$ & $2.2-2.4$ & $2.4-2.6$ & $2.6-5.0$\\
 \hline
 $1.87-2.0$ & $2.0\times 10^{-12}$  & $2.8\times 10^{-13}$  & $8.5\times 10^{-14}$  &  $2.3\times 10^{-14}$ & $1.2\times 10^{-15}$ \\
 \phantom{0}$2.0-2.2$ & - &  $1.1\times 10^{-12}$  &  $8.3\times 10^{-14}$ &  $2.3\times 10^{-14}$ & $1.2\times 10^{-15}$ \\
 \phantom{0}$2.2-2.4$ & - & - & $2.9\times 10^{-13}$ & $6.9\times 10^{-15}$ & $3.8\times 10^{-16}$ \\
 \phantom{0}$2.4-2.6$ & - & - & - & $9.6\times 10^{-14}$ & $1.0\times 10^{-16}$ \\
 \phantom{0}$2.6-5.0$ & - &  -  & - &  -  & $3.9\times 10^{-16}$ \\
 \end{tabular}
\label{sec:appendix:table:covar}
\end{table}

\clearpage

\addcontentsline{toc}{section}{References}
\bibliographystyle{LHCb}
\bibliography{main,standard,LHCb-PAPER,LHCb-DP}

\newpage
\centerline
{\large\bf LHCb collaboration}
\begin
{flushleft}
\small
R.~Aaij$^{31}$,
C.~Abell{\'a}n~Beteta$^{49}$,
T.~Ackernley$^{59}$,
B.~Adeva$^{45}$,
M.~Adinolfi$^{53}$,
H.~Afsharnia$^{9}$,
C.A.~Aidala$^{79}$,
S.~Aiola$^{25}$,
Z.~Ajaltouni$^{9}$,
S.~Akar$^{64}$
% ...

{\footnotesize \it

% ...
$ ^{9}$Universit{\'e} Clermont Auvergne, CNRS/IN2P3, LPC, Clermont-Ferrand, France\\
% ...
$ ^{25}$INFN Sezione di Milano, Milano, Italy\\
% ...
$ ^{31}$Nikhef National Institute for Subatomic Physics, Amsterdam, Netherlands\\
% ...
$ ^{45}$Instituto Galego de F{\'\i}sica de Altas Enerx{\'\i}as (IGFAE), Universidade de Santiago de Compostela, Santiago de Compostela, Spain\\
% ...
$ ^{49}$Physik-Institut, Universit{\"a}t Z{\"u}rich, Z{\"u}rich, Switzerland\\
% ...
$ ^{53}$H.H. Wills Physics Laboratory, University of Bristol, Bristol, United Kingdom\\
% ...
$ ^{59}$Oliver Lodge Laboratory, University of Liverpool, Liverpool, United Kingdom\\
% ...
$ ^{64}$University of Cincinnati, Cincinnati, OH, United States\\
% ...
$ ^{67}$Syracuse University, Syracuse, NY, United States\\
% ...
$ ^{79}$University of Michigan, Ann Arbor, United States, associated to $^{67}$\\
\bigskip
% ...
}
\end{flushleft}

\end{document}